\documentclass[pre,aps,twocolumn,,amsmath,amssymb,showpacs,showkeys,floatfix]{revtex4-1}
\usepackage{graphicx}
\usepackage{amsfonts}
\usepackage{epsfig}
\usepackage{dcolumn}
\usepackage{amsthm}
\usepackage{color}
\usepackage{float}
\usepackage{latexsym}
\usepackage{amscd}
\usepackage{hyperref}
\usepackage[mathscr]{eucal}

\usepackage{bm}      
\usepackage{ifpdf}
\usepackage{bbm}

\begin{document}

\newcommand{\newc}{\newcommand}

\newc{\beq}{\begin{equation}}
\newc{\eeq}{\begin{equation}}
\newc{\ovl}{\overline}
\newc{\bc}{\begin{center}}
\newc{\ec}{\end{center}}
\newc{\tr}{\mbox{tr}}
\newc{\pd}{\partial}
\newc{\dqv}{\delta\vec{q}}
\newc{\dpv}{\delta\vec{p}}
 \newc{\f}{\frac}

\title{Signatures of bifurcation on quantum correlations: Case of quantum kicked top}
\author{Udaysinh T. Bhosale}
\email{udaybhosale0786@gmail.com}
\author{M. S. Santhanam}
\email{santh@iiserpune.ac.in}
\affiliation{Indian Institute of Science Education and Research, Dr. Homi Bhabha Road, Pune 411 008, India.}

\date{\today}

\begin{abstract}
Quantum correlations reflect the quantumness of a system and are useful resources for quantum information and 
computational processes. The measures of quantum correlations do not have a classical analog and yet are influenced
by the classical dynamics.  
In this work, by modelling the quantum kicked top as a multi-qubit system, the effect of classical bifurcations on 
the measures of quantum correlations such as quantum discord, geometric discord, Meyer and Wallach $Q$ measure is 
studied. The quantum correlation measures change rapidly in the vicinity of a classical bifurcation
point. If the classical system is largely chaotic, time averages of the correlation measures are in good
agreement with the values obtained by considering the appropriate random matrix ensembles. The quantum correlations
scale with the total spin of the system, representing its semiclassical limit. In the vicinity of the 
trivial fixed points of the kicked top, scaling function decays as a power-law. In the chaotic limit, for large
total spin, quantum correlations saturate to a constant, which we obtain analytically, based on random
matrix theory, for the $Q$ measure. We also suggest that it can have experimental consequences.

 \end{abstract}

\maketitle

\section{Introduction}
\label{sec:Introduction}

It is well established by more than half a century of quantum chaos research 
that many of the properties of quantum systems can be understood in terms of classical 
objects such as periodic orbits and their stability \cite{StockmannBook}.
 For classically integrable systems, Einstein-Brillouin-Keller quantization method relates
the quantum spectra and the classical action \cite{Stone05} while for chaotic systems 
Gutzwiller's trace formula represents such an approach
connecting the quantum spectra and the classical periodic orbits \cite{GutzwillerBook}.
The advent of quantum information and computation has opened up newer
scenarios in which novel quantum correlations did not have corresponding classical 
analogues. Quantum entanglement is one such phenomena without a
classical analogue. The von Neumann entropy, a measure of quantum entanglement
for a bipartite pure state, captures correlations with purely quantum origins that
are stronger than classical correlations. A host of such measures are now widely 
used in the quantum information theory to quantify stronger than classical correlations.

Quantum correlations do not have exact classical analogues, yet
they are surprisingly affected by the classical dynamics. For instance, in the context
of chaotic systems, it is known that upon variation of a parameter, as chaos 
increases in the system the entanglement also increases and saturates to a value 
predicted based on random matrix theory \cite{Mehtabook}. Recently, this was experimentally demonstrated 
for an isolated quantum system consisting of three superconducting qubits as a 
realisation of quantum kicked top \cite{Neill2016}. It was shown that larger values of entanglement 
corresponds to regimes of chaotic dynamics \cite{arul01}. Theoretically, not just the chaotic 
dynamics but indeed the structure and details of classical phase space, such as the
presence of elliptic islands in a sea of chaos, is known to affect the entanglement \cite{Santhanam2008}. 

Quantum entanglement is an important resource for quantum information 
processing and computational tasks. However, it does not capture all the
correlations in a quantum system. It is possible for unentangled states 
to display non-classical behaviour implying that there might be residual quantum 
correlations beyond what is measured by entanglement. In addition, it is now known 
that entanglement is not the only ingredient responsible for speed-up in
quantum computing \cite{Bennett1999,Niset2006,Horodecki2005}. For mixed state quantum
computing model, discrete quantum computation with one qubit (DQC1),
experiments have shown that some tasks can be speeded up over their classical counterparts even using non-entangled,
i.e., separable states but having non-zero quantum correlations \cite{Knill,Animesh,laflammeoct11}. 
Hence, quantification of 
{\sl all} possible quantum correlations is important. For this purpose, measures like 
quantum discord \cite{Ollivier,vedral} and geometric discord \cite{Dakic,Luo_geometric}, 
Leggett-Garg inequality \cite{LeggettGarg} and a host of others are widely used.

Quantum discord is independent of entanglement and no simple ordering relations
between them is known \cite{Virmani2000,ModiRMP}. 
Entanglement may be larger than
quantum discord even though for separable states entanglement always vanishes but quantum discord may be
nonzero, and thus is less than quantum discord \cite{luo,ARPRau2010,ModiRMP}.
This shows that discord and in general all quantum correlation measures are more fundamental than entanglement
\cite{Maziero2009}. It is shown that two-qubit quantum discord in a dissipative dynamics under Markovian 
environments vanishes only in the asymptotic limit where entanglement suddenly disappears \cite{Werlang2009}. 
Thus, the quantum algorithms that make use of quantum correlations, represented in discord, might be more robust 
than those based on entanglement \cite{Werlang2009}. 
This shows that studying quantum correlation, in general, in a given system is important from the point of view 
of decoherence which is inevitably present in almost all experimental setups.

In the last decade, many experimental and theoretical studies of discord were
performed \cite{Celeri2011}. A recent experiment realizes quantum advantage with zero entanglement 
but with non-zero quantum discord using a single photon's polarization and its path as two qubits
\cite{Maldonado2016}. Other experiments have estimated the discord in an anti-ferromagnetic
Heisenberg compound \cite{Mitra2015} and in Bell-diagonal states \cite{Moreva2016}.
In the context of chaotic systems, e.g., the quantum kicked top, the dynamics of discord reveals the
classical phase space structure \cite{VaibhavMadhok2015}.
In this paper, we show that period doubling bifurcation \cite{ChaosForEngineers} in the 
kicked top leaves its signature in the dynamics of quantum correlation measures such as
discord and geometric discord, including the multipartite entanglement measure Meyer and 
Wallach $Q$ measure \cite{Meyer02}.

The structure of the paper is as follows: In Sec.~\ref{sec:Background} the measures of quantum correlations
used are introduced. In Sec.~\ref{sec:KickedTop} the kicked top model is introduced. In Sec.~\ref{sec:Bifurcation} 
results on the effects of the bifurcation on the time averages of these measures of quantum correlations are given. 
In Sec.~\ref{sec:QuantumChaos} these results are compared with a suitable random matrix model.
In Sec.~\ref{sec:Scaling} scaling of these time averaged measures is studied as a function of total spin.

\section{Measure of quantum correlations}
\label{sec:Background}
\subsection{Quantum Discord}

Quantum discord is a measure of all possible quantum correlations including and beyond entanglement in a quantum 
state. In this approach one removes the classical correlations from the total correlations of the system.
For a bipartite quantum system, its two parts labelled $A$ and $B$,
and represented by its density matrix $\rho_{AB}$,
if the von Neumann entropy is ${\mathcal H}(\rho_{AB}) = -\mbox{Tr} ~ (\rho_{AB} \log \rho_{AB})$, then the
total correlations is quantified by the quantum mutual information as,
\begin{eqnarray}
{\mathcal I}(B:A)  &=& {\mathcal H}(B) + {\mathcal H}(A) - {\mathcal H}(B,A) \label{eq5}.
\end{eqnarray}

In classical information theory, the mutual information based on Baye's rule is given by
\begin{eqnarray}
I(B:A) &=& H(B) - H(B|A)  
\label{eq6}
\end{eqnarray}
where $H(B)$ is the Shannon entropy of $B$. The conditional entropy $H(B|A)$ is the average of the Shannon entropies 
of system $B$ conditioned on the values of $A$. It can be interpreted as the ignorance of $B$ given the information 
about $A$.

Quantum measurements on subsystem $A$ are represented by a positive-operator valued measure
(POVM) set $\{\Pi_{i}^{}\}$, such that the conditioned state of $B$ given outcome $i$ is
\begin{equation}
 \rho_{B|i}=\mbox{Tr}_{A}(\Pi_i \rho_{AB})/p_i\;\;\mbox{and}\;\; p_i=\mbox{Tr}_{A,B}(\Pi_i \rho_{AB})
\end{equation}
and its entropy is $\tilde{\mathcal H}_{\{\Pi_i\}} (B|A)=\sum_i p_i {\mathcal H}(\rho_{B|i})$.
In this case, the quantum mutual information is
${\mathcal J}_{\{\Pi_i\}} (B:A)= {\mathcal H}(B)- \tilde{\mathcal H}_{\{\Pi_i\}} (B|A) $.
Maximizing this over the measurement sets $\{\Pi_i\}$ we get
\begin{eqnarray}
{\mathcal J}(B:A)&=&\mbox{max}_{\{\Pi_i\}}\left({\mathcal H}(B)- \tilde{\mathcal H}_{\{\Pi_i\}} (B|A)\right)\nonumber\\
&=&{\mathcal H}(B)-\tilde{\mathcal H}(B|A)
\end{eqnarray}
where $\tilde{\mathcal H}(B|A)=\mbox{min}_{\{\Pi_i\}}\tilde{\mathcal H}_{\{\Pi_i\}}(B|A)$.
The minimum value is achieved using rank $1$ POVMs since the conditional entropy is concave over the set of 
convex POVMs \cite{animesh_nullity}.
By taking $\{\Pi_i\}$ as rank-1 POVMs, quantum discord is defined as
${\mathcal D}(B:A)={\mathcal I}(B:A)-{\mathcal J}(B:A)$, such that
\begin{eqnarray}
{\mathcal D}(B:A) = {\mathcal H}(A)-{\mathcal H}(B,A)
+\mbox{min}_{\{\Pi_i\}} \tilde{\mathcal H}_{\{\Pi_i\}} (B|A).
\end{eqnarray}
Quantum discord is non-negative for all quantum states 
\cite{ZurekQuantumDiscord2000,Ollivier,animesh_nullity}, and is subadditive \cite{VaibhavMadhok11}.

\subsection{Geometric Discord}

The calculation of discord involves the maximization of $J(A:B)$ by doing measurements on the subsystem $B$,
 which is a hard problem.
A more easily computable form is geometric discord based on a geometric way \cite{Dakic,Luo_geometric}. 
There are no measurements involved in calculating this measure.
 For the special case of two-qubits a closed form expression is 
 given \cite{Dakic}. Dynamics of geometric discord is studied under a common dissipating environment \cite{Huang2016}.
For every quantum state there is a set of postmeasurement classical states, and the geometric discord is 
defined as the distance between the quantum state and the nearest classical state,
\begin{equation}
D^{G}(B|A) = \min_{\chi \in \Omega_0}\|\rho-\chi\|^2 ~,
\end{equation}
where $\Omega_0$ represents the set of classical states, and
$\|X-Y\|^2 = {\rm Tr}[(X-Y)^2]$ is the Hilbert-Schmidt quadratic norm.
Obviously, $D^{G}(B|A)$ is invariant under local unitary transformations.
Explicit and tight lower bound on the geometric discord for an arbitrary
state of a bipartite quantum system $A_{m \times m} \otimes B_{n \times n}$ is available 
\cite{Luo_geometric,Hassan,Rana}. 
Recently discovered  ways to calculate lower bounds on discord for such general states do not require
tomography and, hence, are experimentally realisable \cite{Rana,Hassan}.

Following the formalism of Dakic {\it et al.} \cite{Dakic} analytical expression for the geometric discord for 
two-qubit states is obtained.
The two-qubit density matrix in the Bloch representation is 
\begin{equation}
\rho = \frac{1}{4} \Big( \mathbbm{1} \otimes \mathbbm{1}
     + \sum_{i=1}^{3} x_{i}\sigma_{i}\otimes\mathbbm{1}
     + \sum_{i=1}^{3}y_{i} \mathbbm{1}\otimes\sigma_{i}
     + \sum_{i,j=1}^{3} T_{ij}\sigma_{i}\otimes\sigma_{j} \Big)
\label{bloch}
\end{equation}
where $x_{i}$ and $y_{i}$ represent the Bloch vectors for the two qubits,
and $T_{ij}={\rm Tr}[\rho(\sigma_{i}\otimes\sigma_{j})]$ are the components
of the correlation matrix. The geometric discord for such a state is
\begin{equation}
D^{G}(B|A) = \frac{1}{4}\left(\|x\|^2 + \|T\|^2 - \eta_{\rm max} \right),
\label{dg}
\end{equation}
where $\|T\|^2 = {\rm Tr}[T^{T}T]$,\ and $\eta_{\rm max}$ is the largest eigenvalue of 
$\vec{x}\vec{x}^{T} + TT^{T}$, whose explicit form is in \cite{Girolami2012}. 

\subsection{Meyer and Wallach $Q$ measure}
In this work, the effects of bifurcation on multipartite entanglement is also studied using the Meyer and Wallach 
$Q$ measure \cite{Meyer02}. This was used to study the multipartite entanglement in spin Hamiltonians 
\cite{ArulLakshminarayan2005,Karthik07,Brown08} and system of spin-boson \cite{Lambert2005}.
The geometric multipartite entanglement measure $Q$ is shown to be simply related to one-qubit 
purities \cite{Brennen}. Making its calculation and interpretation is straightforward.  
If $\rho_i$ is the reduced density matrix of the $i$th spin obtained by tracing out the rest of the spins in a 
$N$ qubit pure state then
 \begin{equation}
 Q(\psi)=2 \left( 1-\frac{1}{N}\sum_{i=1}^{N}\mbox{Tr}(\rho_i^2) \right).
 \end{equation}
This relation between $Q$ and the single spin reduced density matrix purities has led to a generalization of
this measure to multiqudit states and for various other bipartite splits of the chain \cite{Scott2004}.

\section{kicked top}
\label{sec:KickedTop}

The quantum kicked top is characterized by an angular momentum vector ${\bf J} = ( J_x, J_y, J_z )$,
whose components obey the standard angular momentum algebra. Here, the Planck's constant is set to unity.
The dynamics of the top is governed by the Hamiltonian \cite{Haakebook}:
\begin{equation}
H(t) = p J_y + \frac{k}{2j} J_{z}^{2} \sum_{n=-\infty}^{+\infty} \delta (t-n).
\label{eq:single}
\end{equation}
The first term represents the free precession of the top around $y-$axis with angular
frequency $p$, and the second term is periodic $\delta$-kicks applied to the top.
Each kick results in a torsion about the $z-$axis by an angle $(k/2j) ~J_z$.
The classical limit of Eq.~(\ref{eq:single}) is integrable for $k=0$ and becomes
increasingly chaotic for $k > 0$.
The period-1 Floquet operator corresponding to Hamiltonian in Eq.~(\ref{eq:single}) is given by
\begin{equation}
U=\exp\left(-i\frac{k}{2j} J_{z}^{2}\right) \exp\left(-i p J_{y}\right). 
\label{FloquetOperator}
\end{equation}
The dimension of the Hilbert space is $2j+1$ so that dynamics can be explored without 
truncating the Hilbert space.
Kicked top was realized in experiments \cite{Chaudhury09} and the range of parameters
used in this work makes it experimentally feasible.

The quantum kicked top for given angular momentum $j$ can be regarded as a quantum 
simulation of a collection of $N=2j$ qubits (spin-half particles)
whose evolution is restricted to the symmetric subspace under the exchange of particles.
The state vector is restricted to a symmetric subspace spanned by
the basis states $\{|j,m\rangle ; (m = -j,-j + 1, . . . ,j )\}$ where $j=N/2$.
It is thus a multiqubit system whose collective behavior is governed by the Hamiltonian in 
Eq.~\ref{eq:single} and quantum correlations between any two qubits can be studied.
Kicked top has served as a useful model to study entanglement 
\cite{Lombardi2011,ShohiniGhose2008, Miller99,arul01,bandyo03,BandyoArulPRE} and its relation to classical 
dynamics \cite{Ghikas2007}.

The classical phase space shown in Fig.~\ref{fig:phasespaceplot1} is a function of coordinates $\theta$ and $\phi$.
In order to explore quantum dynamics in kicked top, we construct spin-coherent states
\cite{Hakke87,Arecchi1972,Glauber1976,PuriBook} pointing along the direction of
$\theta_0$ and $\phi_0$ and evolve it under the action of Floquet operator.
The quantum correlations reported in this paper represent time averages obtained from time evolved spin-coherent 
state.
\begin{figure}[t]
\begin{center}
\includegraphics*[scale=0.345]{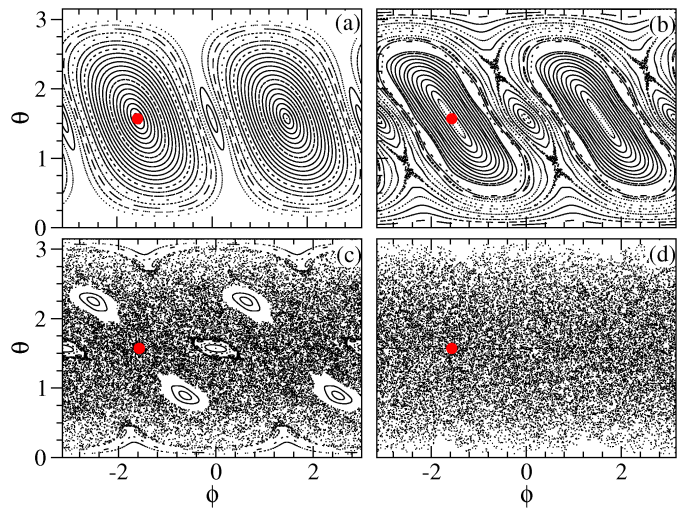}
\caption{(Color online) Phase-space pictures of the classical kicked top for $p=\pi/2$ and
(a) $k=1$, (b) $k=2$, (c) $k=3$ and (d) $k=6$. Red solid circles indicates initial position of the spin 
coherent state.}
\label{fig:phasespaceplot1}
\end{center}
\end{figure}

The classical map for the kicked top is \cite{Haakebook,Hakke87},
\begin{subequations}
\begin{eqnarray}
X^{\prime} &=& (X \cos p + z \sin p) \cos\left(k\left(z\cos p - X \sin p\right)\right)\nonumber\\ 
&& -Y \sin\left(k\left(z \cos p- X \sin p\right)\right)\\
Y^{\prime} &=& (X \cos p + Z \sin p)\sin \left(k\left(Z \cos p-X \sin p\right)\right) \nonumber\\ 
&& +Y \cos\left(k\left(Z \cos p-X\sin p\right)\right)\\
Z^{\prime} &=& -X \sin p + Z \cos p.
\end{eqnarray} 
\label{eq:ClassicalMap}
\end{subequations}
Since the dynamical variables $(X,Y,Z)$ are restricted to the unit
sphere i.e. $X^2+Y^2+Z^2=1$, they can be parameterized in terms 
of the polar angle $\theta$ and the
azimuthal angle $\phi$ as $X = \sin \theta \cos \phi$, $Y = \sin \theta \sin \phi$ and $Z = \cos \theta$. 
We evolve the map in Eq.~(\ref{eq:ClassicalMap}) and determine the values of $(\theta,\phi)$ using the
inverse relations (not shown here). For $p=\pi/2$ additional symmetry properties leads to
a simpler classical map, a case studied in detail in ref. \cite{BandyoArulPRE,VaibhavMadhok2015}.
In this paper two cases namely $p=\pi/2$ and $p=1.7$ are studied which are different from random matrix theory
point of view as explained in Sec.~\ref{sec:QuantumChaos}.
\begin{figure}[t]
\begin{center}     
\includegraphics*[scale=0.34]{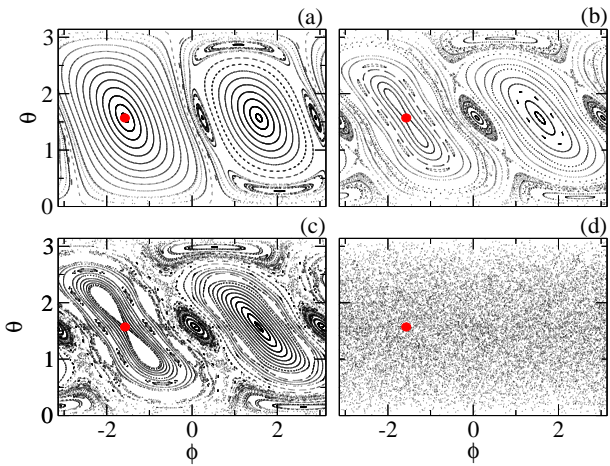}
\caption{(Color online) Phase-space of the classical kicked top for $p=1.7$ and
(a) $k=1$, (b) $k=1.9$, (c) $k=2.1$ and (d) $k=6$. Red solid circles indicates initial position of the spin 
coherent state.}
\label{fig:phasespaceplot2}
\end{center}
\end{figure} 

\section{Effect of Bifurcation}
\label{sec:Bifurcation}

Firstly, we consider the case of $p=\pi/2$. If kick strength is $k=1$, then the phase space
is largely dominated by invariant tori as seen in Fig.~\ref{fig:phasespaceplot1}(a). In particular,
the trivial fixed points of the map at $(\theta,\phi)=(\pi/2,\pm\pi/2)$ visible in 
Figs.~\ref{fig:phasespaceplot1}(a,b) become unstable at $k=2$. As $k$ increases further, the
new fixed points born at $k=2$ move away (see Fig.~\ref{fig:phasespaceplot1}(c)). For $k=6$, the
phase space is largely chaotic with no islands visible in \ref{fig:phasespaceplot1}(d)).
In kicked top, the period doubling bifurcation is the route for regular to chaotic transition.

Second case that is studied here is $p=1.7$.
As seen in Fig.~\ref{fig:phasespaceplot2}(a-d), the phase space displays similar features as 
in the case of $p=\pi/2$ except that the trivial fixed point $(\theta,\phi)=(\pi/2,-\pi/2)$
now loses stability at numerically determined $k=1.76$ while $(\theta,\phi)=(\pi/2,\pi/2)$ loses at $k=2.2$. 
The dark circle, marking the point $(\theta_0,\phi_0)=(\pi/2,-\pi/2)$ in Figs.~\ref{fig:phasespaceplot1}
and \ref{fig:phasespaceplot2}, is the initial position of the spin-coherent state wavepacket.

To study the effect of bifurcation on the quantum correlation and 
multipartite entanglement measures, multiqubit representation of the system is used. For particular value of $j$ the 
system can be decomposed into $N=2j$ qubits. The reduced density matrix of two qubits is calculated by tracing 
out all other $N-2$ qubits \cite{WangMolmer2002,HuMing2008} after every application of the Floquet 
map. We use the reduced density matrix to compute the various measures of correlation.
As all the qubits are identical, the correlations measures do not depend on the actual 
choice of two qubits. Similarly, while calculating $Q$ measure one needs to compute reduced density matrix
of only one qubit. 

The spin-coherent state at time $t=0$ denoted as $|\psi(0)\rangle$ 
is placed at 
the fixed point $(\theta,\phi)=(\pi/2,-\pi/2)$ (red solid circle in Figs.~\ref{fig:phasespaceplot1} 
and \ref{fig:phasespaceplot2}) undergoing a period doubling bifurcation. The state 
$|\psi(0)\rangle$ is evolved by the Floquet operator $\hat{U}$ as 
$|\psi(n)\rangle= U_{}^{n} |\psi(0)\rangle$. We apply the numerical iteration scheme given in 
refs. \cite{Miller99,AsherPeres1996} for time evolving the initial state.
At every time step, discord $D$, geometric discord $D^G$ and, Meyer and Wallach $Q$
measure is calculated for given value of $k$.  The results shown in Figs.~\ref{fig:TimeEvolveDiscord1}
and \ref{fig:TimeEvolveDiscord2} represent time averaged values of $D$, $D^G$ and $Q$ for every $k$.
\begin{figure}[t]
\begin{center}     
\includegraphics*[scale=0.345]{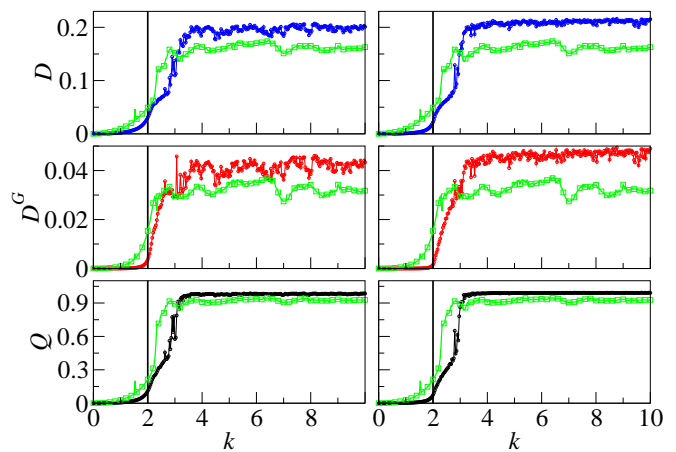}
\caption{(Color online) 
Average discord, geometric discord and $Q$ measure as a function of $k$ for $p=\pi/2$. Left (right) column is for 
$j=50$ ($j=120$). 
For comparison purposes, $j=10$ case is shown in every graph as square (green) symbols. 
The vertical line marks the position of bifurcation at $k=2$.  
}
\label{fig:TimeEvolveDiscord1} 
\end{center}
\end{figure} 

For both cases of $p=\pi/2$ (Fig.~\ref{fig:TimeEvolveDiscord1}) and $p=1.7$ (Fig.~\ref{fig:TimeEvolveDiscord2}), 
the results are shown for two different values of $j$, namely, $j=50$ and $j=120$. For comparison,
the case of $j=10$ qubits is also shown in Fig.~\ref{fig:TimeEvolveDiscord1}. Broadly, in all the 
cases, the quantum correlation measures $D$, $D^G$ and $Q$ respond to the classical bifurcation 
in a similar manner; by displaying a jump in the mean value from about 0 to a non-zero value. 
This can be understood as follows. When the elliptic islands are large, as is the case when
{$0 < k < 2$ for $p=\pi/2$ and $0 < k < 1.76$ for $p=1.7$ case,}
the evolution of the spin-coherent state placed initially at $(\theta,\phi)=(\pi/2,-\pi/2)$ 
is largely confined to the same elliptic islands. As the bifurcation point is approached, the 
local instability in the vicinity of the fixed point evolves part of the coherent state into the 
chaotic layers of phase space. This leads to an increase in the values of correlation measures. 
Note that increasing chaos leads to an increase in entanglement too. When $j$ is 
increased, the width of coherent state $\sigma \propto 1/\sqrt{j}$ 
becomes narrower and closely mimics the classical evolution \cite{Hakke87}. Thus, as $j$ increases, we expect the
quantum correlations to sharply respond to classical bifurcation at $k=2$. Indeed, as seen in
Fig.~\ref{fig:TimeEvolveDiscord1}, the quantum correlations changes sharply at $k=2$
for $=120$ in comparison with the case of $j=10$. To understand the details of Fig.~\ref{fig:TimeEvolveDiscord1}
consider two values of $j$, e.g., $j=j_1$ and $j=j_2$ such that $j_2 > j_1$.
The slow decay of $\sigma$ as $j \to \infty$ implies that the response of
quantum correlations to classical bifurcation becomes perceptible only when $ |j_2 - j_1| >> 1$.
Thus, relative changes are easily seen when quantum correlations for $j=120$ is compared 
with $j=10$ case rather than with that of $j=50$.
The approach to semiclassics, $\hbar \to 0$ limit, discussed in Section~\ref{sec:Scaling}
provides a quantitative support to this picture.


\begin{figure}[t]
\begin{center}     
\includegraphics*[scale=0.345]{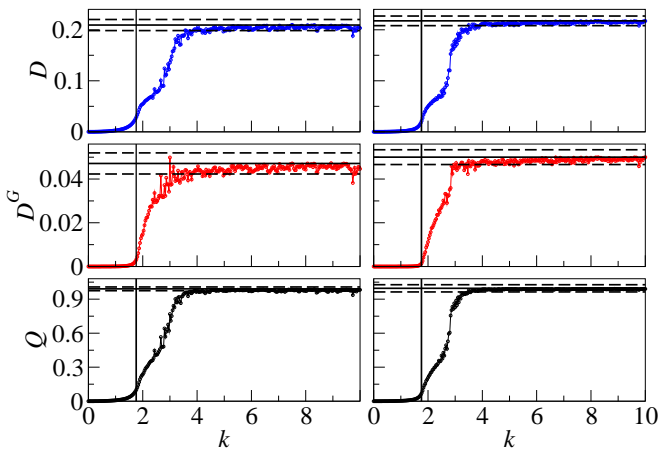}
\caption{(Color online) 
Average discord, geometric discord and $Q$ measure as a function of $k$ for $p=1.7$.
Left (right) column is for $j=50$ ($j=120$). The solid horizontal line represents the long time
average of an initial state from the bifurcation point evolved using the operator $U_{CUE}$.
The dashed line represent the standard deviation from the average value. 
Vertical line marks the position of bifurcation approximately at $k=1.76$.  
}
\label{fig:TimeEvolveDiscord2} 
\end{center}
\end{figure}

\section{Correlation measures and random matrix theory}
\label{sec:QuantumChaos}
Next, we show that the saturated values for $D$, $D^G$ and $Q$ after bifurcation has taken place at $k=k_b$, can be 
obtained from random matrix considerations. The kicked top is time-reversal invariant and as a consequence its 
Floquet operator in the globally chaotic case has the statistical properties of a random matrix chosen from the circular orthogonal ensemble 
(COE) \cite{Kus1988}. For kicked top, the statistical properties of eigenvectors of its Floquet operator are in good 
agreement with COE of random matrix theory \cite{Kus1988}. Apart from time-reversal symmetry, the kicked top 
additionally has the parity symmetry, $\widehat{R}_y=\mbox{exp}(-i\pi j_y)$ that commutes with the Floquet operator 
for all values of $p$. As $\widehat{R}_y^2=I$, the eigenvalues of $\widehat{R}_y$ are $+1$ and $-1$. Thus, in the 
basis of the parity operator, the Floquet operator has a block-diagonal structure consisting of two blocks 
associated with the positive-($+1$) or negative-parity ($-1$) eigenvalues. Thus, due to the parity symmetry, the 
kicked top is statistically equivalent to a block-diagonal random matrix (block diagonal in the basis in which the 
parity operator is diagonal) whose blocks (corresponding to the eigenvalues $\pm 1$) are sampled from the COE 
\cite{Mehtabook}. 
If $p=\pi/2$ the kicked top posseses additional symmetries \cite{Kus1988}, the case which is not considered in this
section. In this section, the case when $p=1.7$ is studied in detail.

Firstly, a block-diagonal COE as the appropriate ensemble of random matrices for modeling the kicked-top Hamiltonian 
is used. Since the basis here is that of eigenvectors of the parity operator $\widehat{R}_y$, this matrix is then 
written in the $|j,m\rangle$ basis. Finally, this matrix is used to evolve the coherent state and compared with the evolution 
done using the Floquet operator in the globally chaotic case ($k=10$). 
The results are presented in Fig.~\ref{tevol1} and summarised in Table \ref{table1}.
\begin{figure}[t]
\begin{center}     
\includegraphics*[scale=0.335]{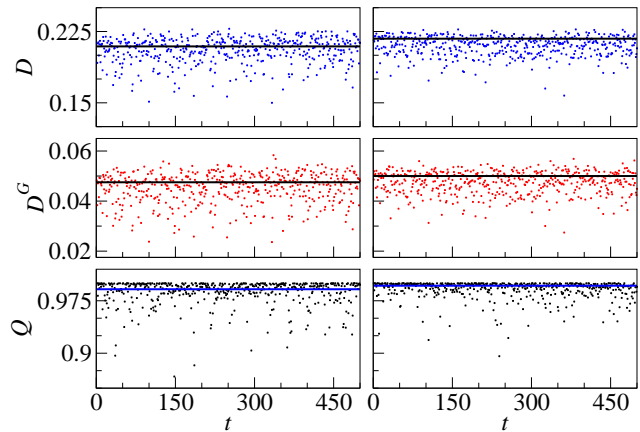}
\caption{(Color online)
Time variation of the correlation measures using kicked-top Floquet operator for $j=50$ (left) and for $j=120$ 
(right) for the globally chaotic case ($k=10$ and $p=1.7$). Horizontal line corresponds to time average of the correlation measures using a
COE matrix of the respective case. 
}
\label{tevol1}
\end{center}
\end{figure}

\begin{figure*}[t]
\begin{center}
\includegraphics*[width=5.6in]{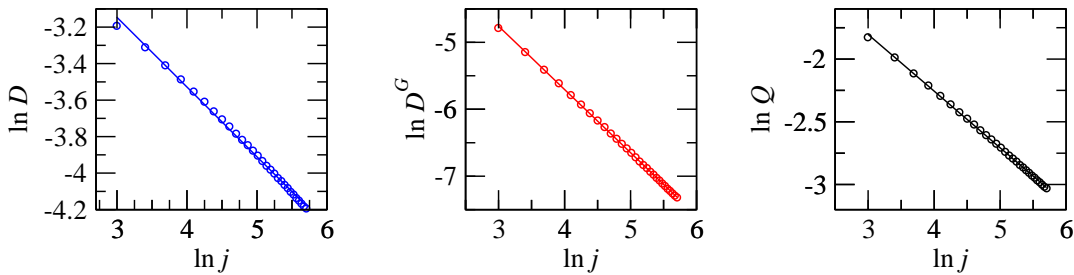}
\end{center}
\caption{(Color online) The variation of time-averaged quantum correlations (circles) as a function of $j$. 
The lines are the power law fits given in Eq.~(\ref{eq:linearFit}).
}
\label{hbarscaling1}
\end{figure*}

Fig.~\ref{tevol1} shows the evolution of $2$-qubit discord, geometric discord and Meyer-Wallach $Q$ measure for 
$j=50$ and $j=120$ when acted by kicked-top Floquet operator with $k=10$.
At this kick strength the classical phase space of kicked
top is largely chaotic with no visible regular regions.
As Fig.~\ref{tevol1} and Table \ref{table1} reveal, the dynamics of 
various correlations measures under the action of COE matrix is similar to that
of kicked-top Floquet operator in its chaotic regime with $k=10$.  
%
%
While this is not entirely unexpected,
the values of the three measures listed in Table \ref{table1} closely agree with those
obtained after bifurcation takes place at $k=k_b$, but at values of kick strengths much
less than $10$ considered in Fig.~\ref{tevol1}. 
Time averages listed in Table \ref{table1} are plotted in Fig.~\ref{fig:TimeEvolveDiscord2}
along with the standard deviation of the individual measures. It can be seen that the agreement
between these values and that of Floquet operator begins to emerge at around $k=4$ which is much less than
$k=10$. The position of the coherent state in this case is $(\theta,\phi)=(\pi/2,-\pi/2)$. It should be noted 
that in the globally chaotic case these results are independent of the initial position of the coherent state.

{
It can be seen from Table \ref{table1} and Fig.~\ref{fig:TimeEvolveDiscord2} that the time averages 
of quantum correlations for the kicked top are systematically, although slightly, lower than that predicted by 
the circular orthogonal ensemble (COE) of random matrix theory. The agreement improves as $j \to \infty$.
Hence, these deviations can be attributed to finite $j$ effect.
Similar systematic deviations from RMT were
observed in the study of the log-negativity in kicked rotor system \cite{Uday12}. In this case too,
the deviations decreased as the corresponding Hilbert space dimensions were increased.
}

\begin{table}[ht]
\centering
\begin{tabular}{|c| c| c|| c| c|}
\hline \hline 
         &   $j=50$ & $j=50$ & $j=120$ & $j=120$ \\
Measure & Floquet & COE & Floquet & COE \\
\hline 
Discord           & $0.205$  & $0.209$  & $0.217$  & $0.217$ \\
Geometric discord & $0.045$  & $0.047$  & $0.049$  & $0.050$  \\
$Q$ measure       & $0.986$  & $0.991$  & $0.994$  & $0.996$  \\
\hline
\end{tabular}
\caption{Mean value of correlation measures averaged over $1000$ time steps of
evolution of a coherent state with the Floquet matrix (with $k=10$) and the COE matrix.
The COE values are represented in Fig.~\ref{fig:TimeEvolveDiscord2} as horizontal lines.}
 \label{table1}
 \end{table}

\section{Scaling with Planck volume}
\label{sec:Scaling}
Kicked top is a finite dimensional quantum system and the volume of its 
Planck cell is $V=4\pi/(2j+1)$. For large $j$, $V \propto 1/j$.
It is natural to ask how the measures of quantum correlation scale with this volume when kick 
strength corresponds to $k=k_b$ where $k_b$ is a bifurcation point. In Fig.~\ref{hbarscaling1}, we show the 
variation in the time average of $D$, $D^G$ and $Q$ as a
function of $j$ for $k=k_b$. Here, $k_b=2$ and $p=\pi/2$. The coherent state is placed at the corresponding trivial
fixed point $(\theta,\phi)=(\pi/2,-\pi/2)$ and the time average is taken over $500$ steps.
For ~$j >>1$, the correlation measures scale with $j$ approximately in a power-law of the form 
$j^{-\mu}$, $\mu$ is the scaling exponent. The power law fits through linear regression for 
the numerically computed correlations measures shown in Fig.~\ref{hbarscaling1} 
are consistent with
\begin{equation}
 D \propto j^{-\mu_1},\;\;\; D^G\propto j^{-\mu_2}\;\;\;\mbox{and}\;\;\;Q \propto j^{-\mu_3},
\label{eq:linearFit}  
\end{equation}
where $\mu_1=0.382\pm 0.003 $,  $\mu_2=0.944$  and $\mu_3=0.451$. 
The uncertainty values are estimated by numerical linear regression
The uncertainties in the estimates
for $\mu_2$ and $\mu_3$ are of the order of $10^{-8}$ and hence negligible.
Identical power-law scaling is obtained for the other trivial fixed point at $(\theta,\phi)=(\pi/2,\pi/2)$ 
with exponents $\mu_i$ approximately same as given in Eq.~(\ref{eq:linearFit}).
The quantum correlations tend to zero as $V \to 0$ ($j\to\infty$) indicating that for any finite $j$ quantum
correlations, however small it might be, would continue to exist. 
As the wavepacket becomes more 'classical' and the underlying dynamics is regular, we expect the
quantum correlations to decrease with $j$. This is another indication that the regular regions
in the vicinity of the fixed point undergoing bifurcations
affect the quantum correlations deep in the semiclassical regime.

The appearance of power-law scaling can be understood
for the case $k=2$ when the regular region is large and the chaotic layer is a tiny fraction of the 
entire phase space. The presence of chaotic layer has a strong influence on quantum correlations. 
Note that for $j>>1$, the width of
the spin-coherent state $\sigma =  j^{-1/2}$ becomes small and its evolution is mostly confined to the large elliptic islands
in Fig.~\ref{fig:phasespaceplot1}(a,b). 
As a result, it can be argued that the strength of the overlap of coherent state with the chaotic
layer is indicative of quantum correlations. 
Since $\sigma = j^{-1/2}$, for $j>>1$, this overlap 
is small. The slow power-law decay of $\sigma$ might possibly be the reason for similar decay
of quantum correlations as well, as shown in Eq.~(\ref{eq:linearFit}). Since
quantum correlations are affected by the local phase space features,
a complete quantitative explanation for power-law scaling
might require a detailed semiclassical analysis.

Next, we consider the case of a coherent state placed initially
at a bifurcation point leading to a period-2 cycle. The origin of this bifurcation point
is as follows.
The trivial fixed points at $(\theta,\phi)=(\pi/2,\pm\pi/2)$ are easily visible in
Figs.~\ref{fig:phasespaceplot1}(a-b) and \ref{fig:phasespaceplot2}(a-b). If $p=\pi/2$,
these fixed points bifurcate at $k=2$ through a period doubling bifurcation and 
become unstable. In the process, the point $(\theta,\phi)=(\pi/2,\pi/2)$ gives rise to two 
new period-$1$ stable fixed points while 
the point $(\theta,\phi)=(\pi/2,-\pi/2)$ gives rise to a period-2 cycle. For $k > 2$ their 
positions move in the phase space as a function of $k$ and they are stable for $k \leq \sqrt{2}\pi$. 
For $k > \sqrt{2}\pi$, the two fixed points bifurcate into two new period-2 cycles while the period-2 cycle
gives rise to a new period-4 cycle. Their positions for $k=\sqrt{2}\pi$ are shown in
Fig.~\ref{hbarscaling2}. Our interest lies in the fixed point located at $(\theta,\phi)=(\pi/4,0)$.

\begin{figure}[t]
\begin{center}     
\includegraphics*[scale=0.35]{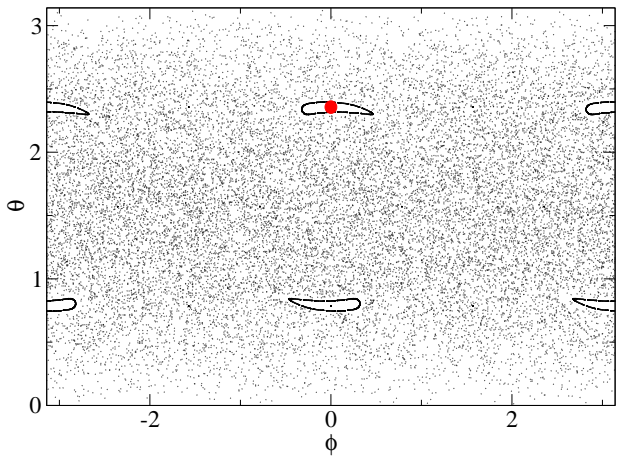}
\caption{(Color online) Phase-space picture of the classical kicked top for $k=\sqrt{2} \pi$.
Red solid circles indicates initial position of the spin coherent state.}
\label{hbarscaling2}
\end{center}
\end{figure}
\begin{figure*}[t]
\begin{center}
\includegraphics*[width=5.6in]{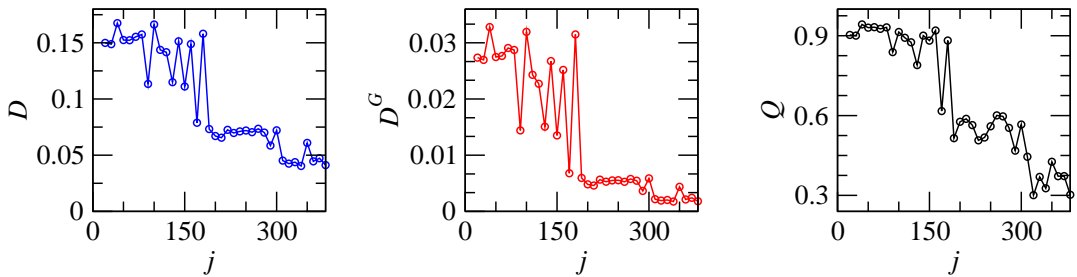}
\end{center}
\caption{(Color online) The variation of time-averaged quantum correlations (circles connected with lines) as a 
function of $j$ for $k=\sqrt{2} \pi$ and initial position of the spin coherent state as shown in 
Fig.~\ref{hbarscaling2}.
}
\label{hbarscaling2M}
\end{figure*}



Fig.~\ref{hbarscaling2M} shows variation of the time average of the quantum 
correlation measures as a function of $j$ for the initial coherent state placed at 
this fixed point. It can be seen that after initial fluctuations the correlations
start to decrease for larger values of $j$.
It should be noted that the area of elliptic islands are
continually shrinking as $k \to \infty$ consistent with the predominance of 
chaotic regions in the phase space. The width of the spin-coherent state $|\psi(0)\rangle$ is equal 
to $1/\sqrt{j}$. For small values of $j$, the width of $|\psi(0)\rangle$ is much larger 
than that of the regular elliptic island as shown in Fig. \ref{hbarscaling2}.
Hence, there is a pronounced overlap of the state $|\psi(0)\rangle$ with the 
chaotic sea. Hence we expect that for small $j$ the quantum correlations will be
reasonably close to their random matrix averages. This is indeed seen in Figs. \ref{hbarscaling2M}
for $1 \le j \le 50$ as the width of $|\psi(0)\rangle$ are at least twice the size of elliptic island. 
For $j>>1$, the width of $|\psi(0)\rangle$ has become much smaller than that of elliptic island. Thus, 
under these conditions we expect smooth decay with increasing $j$, similar to what is seen in 
Fig.~\ref{hbarscaling1}(a-c). Fig.~\ref{hbarscaling2M} do show smooth decay for $j\gtrsim 150$.
Thus, the quantum correlations, on an average, decay as a function of $j$ and the
area of the regular region surrounding the fixed point
undergoing bifurcation strongly affects the quantum correlations.

\begin{figure*}[t]
\begin{center}
\includegraphics*[width=5.73in]{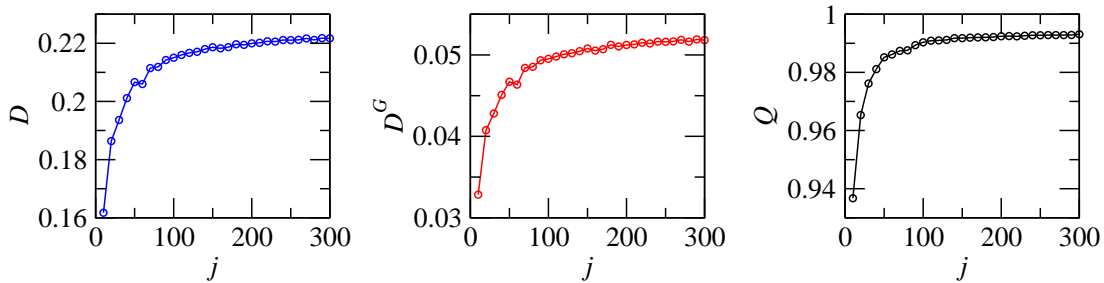}
\end{center}
\caption{(Color online)The variation of time-averaged quantum correlations (circles) as a function of $j$ for the globally chaotic case
($k=10$)..
}
\label{hbarscaling10}
\end{figure*}
\begin{figure}[t]
\begin{center}
\includegraphics*[scale=0.33]{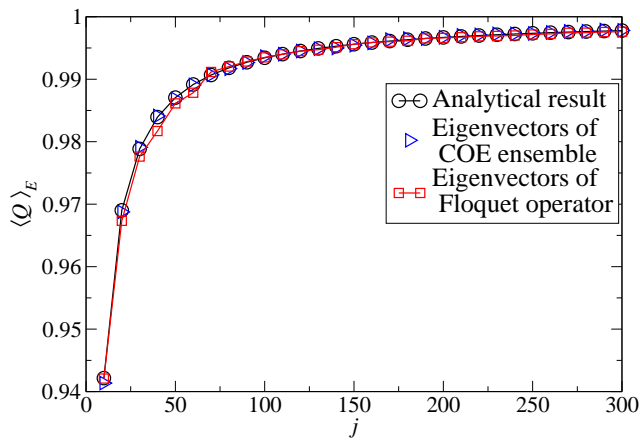}
\caption{(Color online) Average $Q$ measure for the eigenvectors of COE ensemble and that of Floquet operator in 
the globally chaotic case for the parameter range $10\leq k \leq 1000$ and $p=1.7$ are compared with its analytical 
expression given in Eq.~(\ref{eq:AveQRMT}).}
\label{fig:QmeasureRandomRMT}
\end{center}
\end{figure}

Now, we consider kick strength $k=10$ and place the spin-coherent state $|\psi(0)\rangle$
at an arbitrary position in the chaotic sea, namely, $(\theta,\phi)=(1.6707,−1.3707)$. 
Here, the phase space is largely chaotic devoid of any regular regions.
In contrast to the results in Figs. \ref{hbarscaling1} and \ref{hbarscaling2M}, the time averaged 
correlation measures shown in Fig.~\ref{hbarscaling10} increase with $j$.
Based on the results from figs.~\ref{fig:TimeEvolveDiscord2} and \ref{tevol1} we can expect that at every value of 
$j$ the time averaged $D$, $D_G$ and $Q$ agree with those found using apropriate COE ensemble.

For a coherent state the quantum correlation measures are zero.
However, after time evolution, the correlation values will depend on the
corresponding measures for Floquet eigenstates. Thus, it is 
important to study the typical values of these measures for these eigenstates.
This can be analytically obtained for the average $Q$ measure.
An exact analytical formula for the average $Q$ measure is derived (see Appendix~\ref{Appendix1}
for the detailed derivation) for a
typical COE ensemble modelling
the Floquet operator in the globally chaotic case. It is given by
\begin{eqnarray}
\label{eq:AveQRMT}
\langle Q \rangle_E =1-\frac{16j(j+1)}{3(2j+3)(2j+1)^2}. 
\end{eqnarray}
For large $j$, $ \langle Q \rangle_E \approx 1-2/(3j)$ implying that the measure tends to one for large $j$.
The numerically computed correlations for the eigenvectors of COE ensemble and for the eigenvectors of the
Floquet operator under conditions of globally classical chaos are compared with the
analytical result in Eq.~(\ref{eq:AveQRMT}) in Fig. \ref{fig:QmeasureRandomRMT}.


For generating sufficient statistics for the eigenvectors of Floquet operator, we use a range of $k$ values such
that the corresponding classical section does not have any significant regular islands and is highly chaotic.
The analytical result in Eq.~(\ref{eq:AveQRMT}) agrees with that for the eigenvectors of the COE ensemble.
In order to derive similar expressions for the average discord and geometric discord for the eigenvectors of COE,
analytical
expression for the distribution of the matrix elements of the two-qubit density matrix for these class
of states is required. Such an expression is not known yet, to the best of our knowledge. Thus, the derivation of the
average discord and geometric discord as a function of $j$ remains an open question.

It is instructive to compare these results with other well-studied ensembles such as the
Gaussian ensembles. In this case, the states are distributed
uniformly, also known as Haar measure, on the unit sphere. Consider a tripartite random pure state. 
The entanglement between any of its two subsystems shows a transition from being
entangled to separable state as the size of the third subsystem is increased \cite{Uday12,Scott03}.
Another example is that of definite particle states.
This shows algebraic to exponential decay of entanglement when the number of 
particles exceed the size of two subsystems \cite{Vikram11}. For both these cases, discord
and geometric discord between two qubits in a tripartite system
goes to zero as the size of the third subsystem is increased.
It is known that average $Q$ measure for Haar distributed states of $N$ qubits, for large $N$, goes as 
$1-3/2^N$ \cite{Scott03}. In terms of $j(=N/2)$ it equals $1-3/2^{2j}$ implying that the measure tend to $1$ for 
large $j$. But, the rate at which it approaches $1$ is much faster than that for eigenvectors of COE ensemble 
corresponding to the kicked top in globally chaotic case.
In contrast to the standard Gaussian or circular ensemble, the random matrix ensemble
appropriate for the kicked top is COE with additional particle exchange symmetry.
Hence, this ensemble displays different properties from the standard circular or Gaussian
ensembles as far as the quantum correlations are concerned.

Interestingly, it is found numerically in the globally chaotic case that 
$D^G = 0.317~ D - 0.018$
holds good. This is seen in Figs.~\ref{fig:TimeEvolveDiscord1}, \ref{fig:TimeEvolveDiscord2}, \ref{tevol1} and 
\ref{hbarscaling10}. Such simple relation relating $Q$ measure and discord or geometric discord could not be 
discerned. It is known that for two-qubit states, discord and geometric discord are related to each other
by $D^G \geq D^2/2$ \cite{Luo_geometric,girolami,ModiRMP}. This inequality is respected thoughout numerical 
simulations performed here.

\section{Summary and Conclusions}
\label{sec:Summary}
In this paper, we have investigated the effect of classical bifurcations on the measures of quantum correlations 
such as quantum discord, geometric discord and Meyer and Wallach $Q$ measure using kicked top as a model of quantum
chaotic system. In a related work \cite{VaibhavMadhok2015}, signature of classical chaos in the kicked top was 
found in the dynamics of quantum discord and this work explores this relation in the more general context of
quantum correlations including multipartite entanglement. The suitability of kicked top is due to the fact that it can be represented as a collection of 
qubits. Most importantly, this system has been realised in experiments \cite{Chaudhury09}. A prominent feature in its 
phase space is the period-1 fixed point whose bifurcation is associated with the quantum discord climbing from 
nearly $0$ to a value that is in agreement with the numerically determined random matrix equivalent. The transition 
in the quantum discord reflects the qualitative change in the classical phase space; from being dominated by elliptic 
island to a largely chaotic sea with a few small elliptic islands. The other measures we have reported here, namely 
the geometric discord and Meyer and Wallach $Q$ measure, both display similar trends as the quantum discord. 
Other measures of quantum correlations can be expected to display qualitatively similar results.
We have also presented numerical results for the random matrix averages of these quantum correlation measures. 

In general, as
a function of chaos parameter, quantum discord can be expected to increase under the influence of a period-doubling 
bifurcation. However, after the bifurcation has taken place, the saturation to the random matrix average will depend 
on the qualitative nature of dynamics in the larger neighbourhood around the fixed point. It must also be pointed out 
that these results have been obtained through time evolution of a spin-coherent state placed initially on an elliptic
island undergoing bifurcation. For reasonably large elliptic islands, equivalent results could have been obtained by
considering the Floquet states of the kicked top as well.


We have also investigated the fate of quantum correlations in the semiclassical limit as Planck volume tends to 
zero. In the context of kicked top, this limit translates as $j \to \infty$. In the case of bifurcation associated 
with larger islands, as in $k\leq 2$, the measures of quantum correlations
decreases as a function of $j$ and tends to $0$ through a slow, approximately 
power-law decay. In the case of bifurcation associated
with smaller islands and creation of higher order periodic cycles the average decay of quantum
correlations is evident but marked by strong fluctuations. The quantum correlation measures reported here have been 
obtained as that for the time average of an evolving spin-coherent state placed initially at a chosen position in 
phase space. However, we note that if the spin-coherent state is placed instead in the chaotic sea initially, then 
a different behaviour is obtained. As a function of $j$, in this case, the quantum correlation measures increases 
and saturates to a constant value that can be understood based on eigenvectors of appropriate random
matrix ensemble.
Evaluation of exact analytical expression for average $Q$ measure for the eigenvectors of the corresponding 
circular unitary ensemble is carried out and agrees very well that for the eigenvectors of the Floquet operator
in the globally chaotic case.

All the results presented in this work emphasise the special role played by the bifurcations and
the associated regular phase space regions in modifying general expectations for the quantum
correlations based on random matrix equivalents.
These results are important from the experimental point of view as the kicked top was first implemented in a system of 
laser cooled Caesium atoms \cite{Chaudhury09}. Recently this model was implemented using superconducting 
qubits \cite{Neill2016}. Here the time-averaged von Neumann entropy has shown very close resembalance, despite 
presence of the decoherence, with the corresponding classical phase-space structure for given parameter
values \cite{Neill2016}. Hence, the detailed effects of bifurcations presented here should be
amenable to experiments as well. The scaling of quantum correlations with the total spin should also
be observable with less than about ten superconducting qubits.

\begin{acknowledgments}
We are very grateful to acknowledge many discussions with Vaibhav Madhok, T. S. Mahesh, Jayendra Bandyopadhyay
and Arul Lakshminarayan. UTB acknowledges the funding from 
National Post Doctoral Fellowship (NPDF) of DST-SERB, India file No. PDF/2015/00050.

\end{acknowledgments}

\appendix
\section{Exact evaluation of $\langle Q \rangle_E$}
\label{Appendix1}

In this Appendix an exact evaluation of the ensemble average of the Meyer and Wallach $Q$ measure is calculated.
The states in the ensemble have idetical qubits and remains unchanged under qubit exchange.
As explained in Section~\ref{sec:KickedTop} one needs to use symmetric subspace spanned by the basis states 
$\{|j,m\rangle ; (m = -j,-j + 1, . . . ,j )\}$. 
Any pure state $|\phi\rangle$ in this basis is given as
\begin{equation}
|\phi\rangle=\sum_{m=-j}^{j}a_m |j,m\rangle\;\;\mbox{where}\;\;\sum_{m=-j}^{j} |a_m|^2 =1.
\end{equation}

In this case the $Q$ measure is given as follows:
\begin{equation}
 Q=1-\frac{4}{(2j+1)^2}\left(\langle S_z \rangle^2+\langle S_{+}\rangle\langle S_{-}\rangle\right)
\end{equation}
where $S_z$ and $S_{\pm}$ are collective spin operators such that $S_z|j,m\rangle=m|j,m\rangle$ and 
$S_{\pm}|j,m\rangle=\sqrt{(j\mp m)(j\pm m+1)} |j,m \pm 1\rangle$ \cite{HuMing2008}. 
The ensemble average is carried over the states such that they have the statistical properties of the eigenvectors
of the COE ensemble. For the state $|\phi\rangle$ one obtains the following expression for the expectation:
\begin{equation}
 \langle S_z \rangle=\sum_{m=-j}^{j} m |a_m|^2.
\end{equation}
This gives
\begin{eqnarray}
\langle S_z \rangle^2 &=&\sum_{m,n=-j}^{j} m n |a_m|^2 |a_n|^2 \nonumber\\
 &=& \sum_{m=n} m^2 |a_m|^4 + \sum_{m\neq n} m n |a_m|^2 |a_n|^2.                
\end{eqnarray}

Now, an exact RMT ensemble-average is carried out \cite{Ullah,Haakebook}. Firstly, one obtains
\begin{equation}
\label{eq:szsquareaverage}
\langle\langle S_z \rangle^2\rangle_E = \sum_{m=n} m^2 \langle |a_m|^4 \rangle_E + 
\sum_{m\neq n} m n \langle|a_m|^2 |a_n|^2 \rangle_E.
\end{equation}

It should be noted that the first expectation is for a given state $|\phi\rangle$ and second expectation with
subscript $E$ denotes the ensemble average over all $|\phi\rangle$ having statistical properties of COE eigenvectors. 
Using the RMT ensemble averages \cite{Ullah,Haakebook}
\begin{equation}
\label{eq:ensembleAve}
\begin{split} 
\langle |a_m|^4 \rangle_E &= \frac{3}{(2j+1)(2j+3)}, \\
\langle|a_m|^2 |a_n|^2 \rangle_E &= \frac{1}{(2j+1)(2j+3)}
\end{split}
\end{equation}
one obtains
\begin{equation}
\begin{split} 
\langle\langle S_z \rangle^2\rangle_E &= \frac{3}{(2j+1)(2j+3)} \sum_{m=n} m^2  +\\
&\frac{1}{(2j+1)(2j+3)} \sum_{m\neq n} m n.  
\end{split}
\end{equation}

The first summation in the above equation is calculated as follows:
\begin{equation}
 \sum_{m=-j}^{j} m^2 = 2\sum_{m=1}^{j} m^2 = \frac{j(j+1)(2j+1)}{3}.
\end{equation}
The second summation is now calculated. Consider the equality:
\begin{eqnarray}
\left( \sum_{m=-j}^{j} m \right) \left( \sum_{m=-j}^{j} n \right) &=& 0.          
\end{eqnarray}
This gives
\begin{eqnarray}
 \sum_{m,n} m n = \sum_{m=n} m^2+ \sum_{m\neq n} mn=0
\end{eqnarray}
Thus,
\begin{eqnarray}
 \sum_{m\neq n} m n = -\sum_{m=n} m^2 =\frac{-j(j+1)(2j+1)}{3}.
\end{eqnarray}

The ensemble average in Eq.~(\ref{eq:szsquareaverage}) is given as follows:
\begin{equation}
\label{eq:sz2average}
\langle\langle S_z \rangle^2\rangle_E = \frac{2j(j+1)(2j+1)}{3(2j+1)(2j+3)}.
\end{equation}

Considering the average of operators $S_{\pm}$ for the state $|\phi\rangle$
\begin{equation}
 \langle S_{\pm} \rangle = \sum a_m a_{m\pm1}^* \sqrt{(j\mp m)(j\pm m +1)}.
\end{equation}
This gives
\begin{eqnarray}
\begin{split} 
&\langle S_{+}\rangle\langle S_{-}\rangle = \sum_{m,n} a_m a_{m+1}^* a_n a_{n-1}^* \\ 
& \sqrt{(j-m)(j+m+1)(j+n)(j-n+1)}.
\end{split}
\end{eqnarray}

It can be seen that the ensemble average will have nonzero contribution only when $m=n-1$. Thus,
\begin{eqnarray}
\begin{split} 
& \langle \langle S_{+}\rangle\langle S_{-}\rangle \rangle_E = \\
 \sum_{m=-j}^{j-1}& \langle |a_m|^2 |a_{m+1}|^2 \rangle_E (j-m)(j+m+1).
\end{split}
\end{eqnarray}

Using Eq.~(\ref{eq:ensembleAve}) following is obtained:
\begin{eqnarray}
\begin{split} 
& \langle \langle S_{+}\rangle\langle S_{-}\rangle \rangle_E = \\
& \frac{1}{(2j+1)(2j+3)}   \sum_{m=-j}^{j-1} (j-m)(j+m+1).
\end{split}
\end{eqnarray}

Calculating the summation as follows:
\begin{eqnarray}
&&\sum_{m=-j}^{j-1} (j-m)(j+m+1)= \sum_{m=-j}^{j-1} (j^2+j-m^2-m) \nonumber\\
&&=2j(j^2+j)-\sum_{m=-j}^{j-1}m -\sum_{m=-j}^{j-1} m^2 \nonumber\\
&&= 2j(j^2+j)+j+j^2-\frac{j(j+1)(2j+1)}{3}.
\end{eqnarray}
Thus,
\begin{eqnarray}
\label{eq:spsmaverage}
 \langle \langle S_{+}\rangle\langle S_{-}\rangle \rangle_E = \frac{2j(j+1)}{3(2j+3)}. 
\end{eqnarray}

Using Eqs.~(\ref{eq:sz2average}) and (\ref{eq:spsmaverage}) the final expression for the ensemble average of 
$Q$ measure, denoted as $\langle Q \rangle_E $, is given as follows:
\begin{eqnarray}
\label{eq:AveQRMTAppendix}
\langle Q \rangle_E =1-\frac{16j(j+1)}{3(2j+3)(2j+1)^2}. 
\end{eqnarray}
This analytical expression is plotted in Fig.~\ref{fig:QmeasureRandomRMT}.

\bibliography{reference22013,reference22}

\end{document}